# The adsorption of helium atoms on coronene cations

Thomas Kurzthaler,[1] Bilal Rasul,[1] Martin Kuhn,[1] Albrecht Lindinger,[2] Paul Scheier[1,*] and Andrew M. Ellis[3,*]

[1] Institut für Ionenphysik und Angewandte Physik, Universität Innsbruck, Technikerstr. 25, A-6020 Innsbruck, Austria

[2] Institut für Experimentalphysik, Freie Universität Berlin, Arnimallee 14, 14195 Berlin, Germany

[3] Department of Chemistry, University of Leicester, University Road, Leicester, LE1 7RH, UK

Email: Paul.Scheier@uibk.ac.at;  andrew.ellis@le.ac.uk

___

**Abstract**

We report the first experimental study of the attachment of multiple foreign atoms to a cationic polycyclic aromatic hydrocarbon (PAH). The chosen PAH was coronene, $C_{24}H_{12}$, which was added to liquid helium nanodroplets and then subjected to electron bombardment. Using mass spectrometry, coronene cations decorated with helium atoms were clearly seen and the spectrum shows peaks with anomalously high intensities ('magic number' peaks), which represent ion-helium complexes with added stability. The data suggest formation of a rigid helium layer consisting of 38 helium atoms that completely covers both faces of the coronene ion. Additional magic numbers can be seen for the further addition of 3 and 6 helium atoms, which are thought to attach to the edge of the coronene. The observation of magic numbers for the addition of 38 and 44 helium atoms is in good agreement with a recent path integral Monte Carlo prediction for helium atoms on neutral coronene. An understanding of how atoms and molecules attach to PAH ions is important for a number of reasons including the potential role such complexes might play in the chemistry of the interstellar medium.

The adsorption of helium on graphite is a well-studied system which illustrates a range of interesting physics. For adsorption to occur the graphite must be cooled to a low temperature and the combination of the low temperature with the weak binding of helium atoms creates a system where quantum properties are important. Experimental studies have investigated features such as the structures of these layers, their phase transitions and the prospects for two-dimensional superfluidity.[1-5] At low temperatures the helium atoms adhere as a $\sqrt{3} \times \sqrt{3}$ commensurate phase, in which one third of the hexagonal sites above the graphitic plane are occupied by helium atoms. These atoms are locked into these locations by the corrugated surface.[6] A single graphene sheet also forms this commensurate layer but here the corrugations are less pronounced than for graphite and so very little additional energy is needed to induce a phase transition from the solid to a liquid.[7,8]

$C_{60}$ and other fullerenes offer the opportunity to probe helium adsorption on finite and curved graphene surfaces. Individual uncharged fullerene molecules are difficult to study in this way but charged fullerenes can be readily investigated using mass spectrometry. Studies of the decoration of both $C_{60}^+$ and $C_{60}^-$ by helium have been reported, as well as the equivalent ions for $C_{70}$.[9,10] In these experiments the fullerene molecules were first isolated in liquid helium nanodroplets and then subjected to electron bombardment, either at low energy (for anions) or higher energy (for cations). The mass spectra show enhanced abundances for ions with extra stabilities which are often referred to as magic number ions. These data have allowed the 1 × 1 commensurate phases to be identified, where each hexagonal and pentagonal face is occupied by a single helium atom. This becomes possible in the fullerenes, unlike in planar graphite and graphene, because of the curvature of the fullerene surface, which increases the distance between helium atoms when sat on adjacent rings.

Polycyclic aromatic hydrocarbons (PAHs) provide a planar graphite-like surface but of finite size. Calvo has recently carried out a series of calculations on how helium atoms will arrange on three selected PAH cations (pyrene, coronene and circumcoronene), as well as for the benzene cation.[11] These calculations used a variety of computational methods and included searches for classical potential energy minima and, for small numbers of attached helium atoms, allowance for quantum delocalization using path integral molecular dynamics (PIMD). These calculations did not attempt to predict particular magic numbers, but were able to identify commensurate structures, as well as helium atoms bound to the edge of the PAH in a more liquid-like arrangement. The question was raised as to whether it would be possible to see evidence for magic number ions for helium-coated PAH ions through mass spectrometric investigations.

This study reports the first experimental study of the decoration of a PAH cation with helium atoms. We focus here on a single PAH, coronene, and have used mass spectrometry to try and identify magic number ions. The importance of this work goes beyond a comparison with the wetting of graphene and the fullerenes and extends into astrochemistry. In particular, PAH cations have been suggested as potential carriers of diffuse interstellar bands (DIBs), as well as some of the unidentified infrared (UIR) bands. There are several hundred of these bands originating from the interstellar medium, with the DIBs occurring in the visible and near-infrared parts of the spectrum, and almost all are unassigned. The prospect of relatively large molecules with substantial carbon-carbon networks being responsible has recently been boosted by the firm assignment of two DIBs to $C_{60}^{+}$.[12] Considerable suspicion has fallen on cationic polycyclic aromatic hydrocarbons (PAHs) as possible carriers of many of the DIBs, since neutral PAHs are not expected to have electronic transitions in the visible and near-infrared.[13,14] For the UIR bands it seems very likely that both neutral and ionic PAHs make a significant contribution.[15] However, to date no firm assignments of any of these bands to specific PAH cations have been made. Furthermore, in the low temperature conditions of interstellar space, some of the PAH cations might combine with various abundant atoms and molecules to form ion-molecule complexes and it is possible that such complexes might be responsible for some of the DIBs and UIR bands. It is therefore important to understand how atoms and molecules attach to PAH cations and, in the longer term, how this might affect their optical spectra. The current study, which focuses on the attachment of helium atoms, is a first step on this journey.

**Experimental**

Helium nanodroplets were produced by supersonic expansion of helium into a vacuum through a 0.4 μm nozzle. The nozzle was cooled to 9.7 K, giving an expected mean droplet diameter of 44 nm ($10^5$ helium atoms).[16] The nanodroplet expansion was collimated by a skimmer with a 0.8 mm diameter aperture, located 8 mm downstream from the nozzle, and the resulting beam then traversed a pick-up region, where coronene vapor was added. To generate a sufficient vapor pressure the solid coronene was placed in an oven and heated to approximately 120 ºC. The doped droplets then continued through to an ionization region, where they were irradiated by electrons with a kinetic energy of 60 eV. Any cations produced were accelerated into the extraction region of a reflectron time-of-flight mass spectrometer (Tofwerk AG, model HTOF). Further details about the apparatus can be found elsewhere.[17]

The mass spectrum extracted was evaluated by means of custom-designed software known as IsotopeFit.[18] This package is used to deconvolute possible overlapping contributions to particular mass spectral peaks by different sized clusters and isotopologues. The software includes the capability to automatically fit mass peaks and subtract background signals. It explicitly considers isotopic patterns of all ions that are expected to contribute to a given peak.

**Results and Discussion**

A section of the mass spectrum recorded is shown in Figure 1. The range chosen includes the coronene parent ion ($m/z$ 300), whose large signal extends far beyond the scale shown. Larger cluster ions, such as the coronene dimer cation, are also seen but the focus here is on the coronene monomer. A long series of ions derived from complexes between helium atoms and the coronene cation, which we will abbreviate as He$_N$Cor$^+$, are identified in Figure 1. Ions with $N$ up to and exceeding 75 have been identified in the mass spectra.

The ion abundance distribution is mostly smooth and declines with $m/z$. This decline seems to accelerate between $m/z$ 450 and 500 before levelling off again. In addition to the broad changes in ion distribution there are some specific He$_N$Cor$^+$ ion signals which are significantly enhanced relative to the subsequent member in the series, He$_{N+1}$Cor$^+$. In the accepted parlance we describe these ions as 'magic number' ions and attribute their excess abundance to ions with enhanced stabilities. Most often this reflects completion of particularly stable structural arrangements, such as the closure of solvation shells.[19]

A trio of magic number ions are seen at $N$ = 38, 41 and 44. These fall into the region where the aforementioned decline in He$_N$Cor$^+$ abundance is particularly rapid. In order to try and interpret what these magic numbers might reveal, we draw upon recent theoretical work. As mentioned earlier, Calvo has investigated the binding of helium atoms to coronene and other PAH cations.[11] However, the aim of that work was to identify possible structural motifs and to make some general observations about rigid and liquid-like configurations of the helium atoms on the cation. No specific attempt was made to predict the magic number ions that might be observed experimentally.

We therefore turn for assistance to a recent investigation of the adsorption of helium atoms onto the *neutral* coronene molecule by Rodríguez-Cantano *et al*.[20] This work employed both classical and quantum simulations to explore possible structures and thermal fluctuations. Both classical and quantum methods identified a particularly stable structure at $N = 38$, which corresponds to a complete coating of the upper and lower faces of the coronene molecule. The arrangement of the atoms is illustrated in Figure 2(a). There is a helium atom located above the central ring of the coronene and the other 18 He atoms are then obtained by two-dimensional packing around the central helium atom in two hexagonal layers, which reflects the hexagonal symmetry of coronene. In the inner layer the helium atoms do not sit above the centres of the outer rings of the coronene because this would put them too close to the central helium atom (optimum He-He distance ~ 3 Å). The inner layer is therefore displaced towards the outer edges of these rings. The second layer then has the helium atoms beyond the carbon rings at positions between two C-H bonds. There is an equivalent arrangement of helium atoms on the lower face of the aromatic plane. According to the calculations by Rodríguez-Cantano *et al*. this is fairly rigid arrangement for neutral coronene and therefore it is likely to be even more rigid for the coronene cation, given the stronger binding between helium and the PAH through induction forces in the ionic version of the complex.

Although classical and quantum calculations agree on the stability of $N = 38$ in the neutral system, there are significant differences for other magic numbers.[20] For example, in the classical simulations $N = 14$ is predicted to be particularly stable. Here there is a helium atom above all seven $C_6$ rings on both sides of the PAH. However, when quantum effects were taken into account, through the path integral Monte Carlo (PIMC) method, the magic character disappears. In the quantum model the $N = 14$ structure has helium atoms which are much more delocalized than for $N = 38$ and has no significantly enhanced stability relative to its $N = 15$ neighbour. Although we might expect differences between the neutral and coronene ion, in our mass spectrometry experiments we likewise see no signal enhancement for $N = 14$, showing that this ion is not magic.

There are also other subtle differences between the quantum and classical calculations on neutral He$_N$Cor. According to the PIMC calculations, the first solvation shell is full at $N = 44$.[20] Addition of an extra helium atom to He$_{44}$Cor results in 44 compact helium atoms plus one far more delocalized helium atom at a mean distance that puts it well above the first layer. It is therefore possible to add an extra six helium atoms to the $N = 38$ system to complete the first solvation shell and these additional helium atoms must add to the edge of the coronene. The edge-located helium atoms will act as a bridge between the helium layers above and below the coronene plane. The mass spectrum in Figure 1 shows magic signals at $N = 41$ and 44, and thus it seems likely that the latter is evidence for complete closure of the first shell for the coronene cation, assuming a direct analogy between the cationic and neutral systems. Addition of six He atoms to the sides of He$_{38}$Cor$^+$ would retain the full D$_{6h}$ symmetry of both the coronene ion and He$_{38}$Cor$^+$ in the limit of a rigid structure, and this arrangement is shown in Figure 2(b). However, we also see magic character for $N = 41$, suggesting that addition of three helium atoms to He$_{38}$Cor$^+$ yields an ion which is almost as stable as He$_{44}$Cor$^+$. The addition of three He atoms instead of six to form a magic structure was not predicted explicitly by the PIMC calculations on neutral He$_N$Cor. In PIMC calculations on He$_N$Cor$^+$, Calvo predicted that the helium atoms around the edge of the coronene cation form a 'slushy' belt, i.e. not fully liquid but nor as rigid as the helium framework above and below the coronene plane.[9] However, in the calculations by Rodríguez-Cantano et al.[18] on the neutral system no such delocalization was seen in the side belt for the specific case of $N = 44$ (a size not considered in the PIMC calculations by Calvo). It may therefore be more appropriate to consider He$_{41}$Cor$^+$ and He$_{44}$Cor$^+$ as cases where the helium atoms on the side are also relatively rigid. For $N = 41$ a trigonal arrangement (overall D$_{3h}$ symmetry in the rigid limit) of these helium atoms is expected.

**Conclusions**

The coating of a PAH cation, coronene, by helium atoms has been reported for the first time. These ions were made by doping liquid helium nanodroplets with coronene molecules and then subjecting the doped droplets to electron ionization. By using mass spectrometry ions with enhanced abundance (magic number ions) are seen for 38, 41 and 44 attached helium atoms. In the case of $N = 38$ and 44 these fit remarkably well with recent path integral Monte Carlo predictions for particularly stable configurations of *neutral* coronene coated with helium. In the stable $N = 38$ structure the helium provides a complete monolayer coating on both sides of the coronene plane. There are essentially two consecutive hexagonal layers of helium atoms around the central He atom on each plane.

The PIMC calculations on the neutral system predict a complete closing of the first shell when a total of 44 helium atoms are added. Here the extra six atoms occupy sites around the edges of the ring structure. Once again, the experimental observations for the cation seem to fit nicely with this prediction. However, we also see a magic number at $N = 41$, which was not predicted by the PIMC calculations. We speculate that here too a stable and fairly rigid structure is formed, this time with three helium atoms around the edge in trigonal arrangement.

It would be interesting to extend this work by replacing helium atoms with molecules, such as $H_2$, which is clearly of strong astrophysical interest because of its abundance in the interstellar medium. Such experiments are planned in the near future.

**Supplementary material**

See supplementary material for a wider scan of the mass spectrum along with additional peak assignments.

**Acknowledgements**

This work was given financial support by the Austrian Science Fund (FWF) Wien (P26635 and I978). AME is also grateful for a short-term scientific mission in aid of this work funded by an EU COST Action (CM1401; Our Astro-Chemical History).

**Figure captions**

1. Section of the mass spectrum recorded for coronene-doped helium nanodroplets. The blue circles above peaks indicate $He_NCn^+$ ions. Magic number ions are identified for $N$ = 38, 41 and 44 in the expanded view in the upper part of the image. Note that other enhanced signals at smaller values of $N$ are affected by underlying contributions from water clusters and so should not be taken as magic number ions. For further information on these peaks and those that are off-scale in this figure, see the supplementary information.

2. Structures proposed for (a) $He_{38}Cn^+$ and (b) $He_{44}Cn^+$. The structure in (a) is an adaptation of that calculated by Rodríguez-Cantano *et al.* for neutral coronene decorated with helium atoms,[20] and shows helium atom locations (in blue) above the upper face of the coronene (an equivalent set of helium atoms lies below). In (b) six additional helium atoms appear around the edge of the coronene.

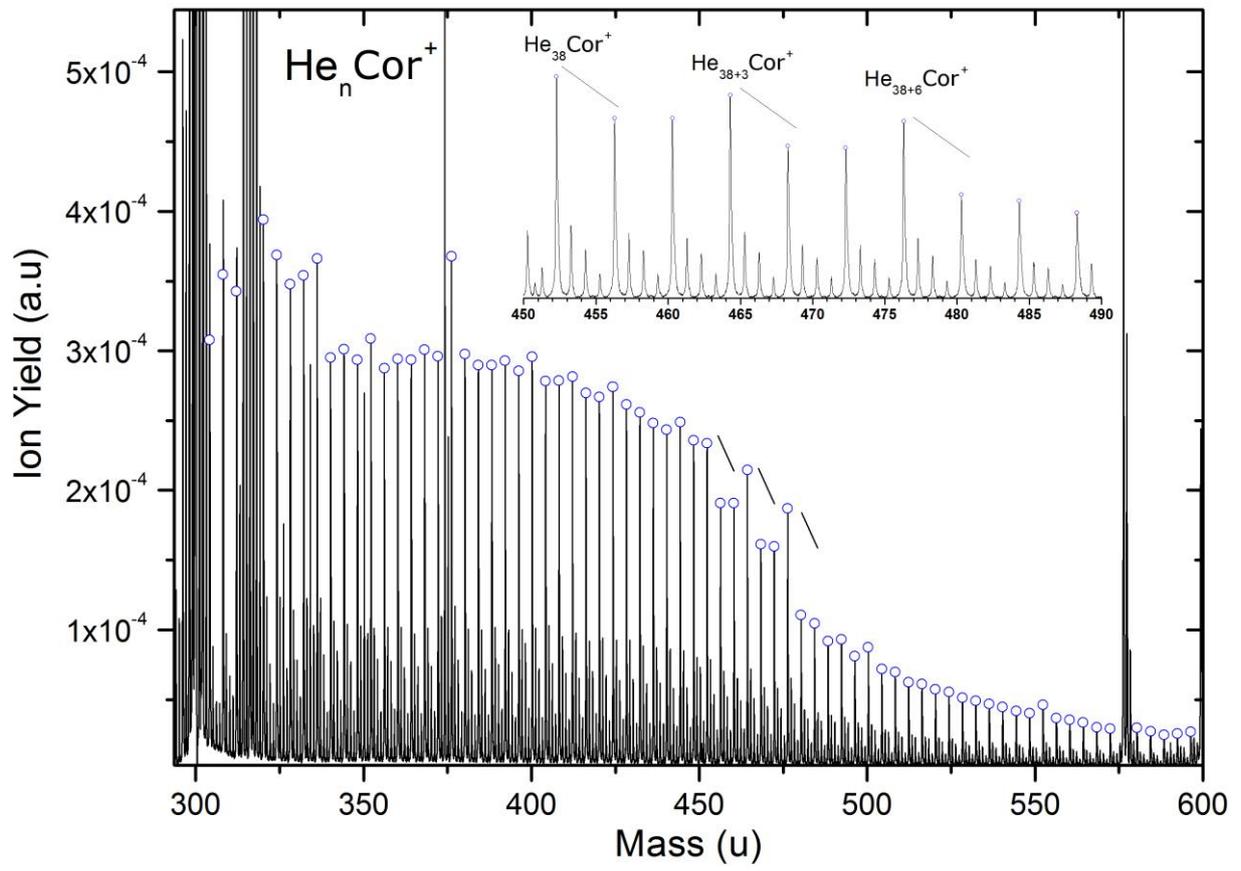

Figure 1

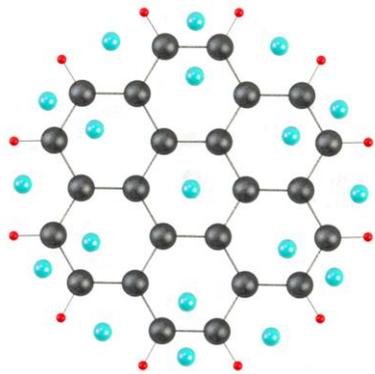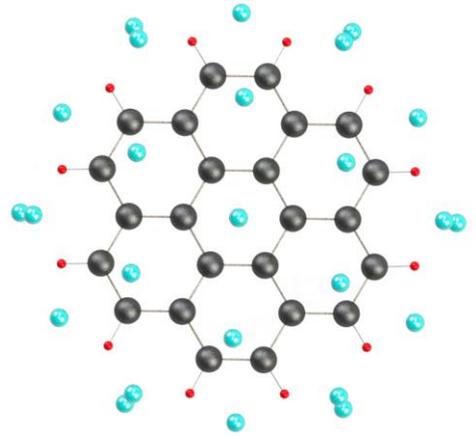

Figure 2